\documentclass[nofootinbib,twocolumn,showpacs,preprintnumbers,amsmath,amssymb,latexsym]{revtex4-1}
\usepackage{latexsym,amssymb,amsfonts,amsmath}
\usepackage{color}
\usepackage[dvips]{graphicx}

\newcommand{\msub}[1]{\ensuremath _{\mbox{\scriptsize #1}}}

\newcommand{\be}{\begin{equation}}
\newcommand{\ee}{\end{equation}}

\newcommand{\is}{I_{s_0}}

\usepackage{graphicx}
\usepackage{dcolumn}
\usepackage{bm}
\usepackage{ulem}

\begin{document}


\title{The localization transition in SU(3) gauge theory}

\author{Tam\'as G.\ Kov\'acs and R\'eka \'A. Vig}
\affiliation{%
Institute for Nuclear Research of the Hungarian Academy of Sciences \\
H-4026 Debrecen, Bem t\'er 18/c, Hungary 
}%




\date{\today}

\begin{abstract}

We study the Anderson-like localization transition in the spectrum of the
Dirac operator of quenched QCD. Above the deconfining transition we determine
the temperature dependence of the mobility edge separating localized and
delocalized eigenmodes in the spectrum. We show that the temperature where the
mobility edge vanishes and localized modes disappear from the spectrum,
coincides with the critical temperature of the deconfining transition. We also
identify topological charge related close to zero modes in the Dirac
spectrum and show that they account for only a small fraction of localized
modes, a fraction that is rapidly falling as the temperature increases.

\end{abstract}

\maketitle

\section{\label{sec:Intro} Introduction}

Quantum Chromodynamics (QCD), the theory of strong interactions, has a rich
structure even at zero chemical potential. At low temperature quarks are
confined in hadrons, while at high enough temperature they are liberated and
form a quark-gluon plasma. The hadronic and plasma phases are not separated by
a genuine phase transition, only a rapid cross-over occurs in between, where
quarks become deconfined \cite{Borsanyi:2010bp}. This deconfining transition
is also accompanied by the restoration of the approximate chiral symmetry that
is spontaneously broken at low temperature.

It is well-known that at low temperature eigenstates of the quark Dirac
operator are delocalized and the corresponding eigenvalues are described by
Wigner-Dyson statistics known from random matrix theory (see
eg.\ \cite{Verbaarschot:2000dy} for a review). Already a long time ago
speculations appeared that the finite temperature chiral transition might be
accompanied or even driven by an Anderson-type localization transition in the
lowest part of the Dirac spectrum \cite{Halasz:1995vd}. If this happens, the
lowest eigenmodes of the Dirac operator become localized and the corresponding
part of the spectrum is described by Poisson statistics.  It was only a decade
later that the first evidence for such a transition was obtained both in an
instanton liquid model \cite{GarciaGarcia:2005vj} and in lattice simulations
\cite{GarciaGarcia:2006gr}. In both cases it was found that when the system
crosses into the high temperature phase, the spectral statistics starts to
move from Wigner-Dyson towards Poisson. However, in the lattice simulation the
used volumes were not large enough to allow a precise quantitative assessment
of the localization transition.

Later on more quantitative lattice studies of the transition appeared both in
QCD and other similar models, and features of an Anderson transition were
confirmed in more detail (see \cite{Giordano:2014qna} for a review). It was
found that at temperatures well above the transition the lowest part of the
Dirac spectrum consists of spatially localized eigenmodes, while eigenmodes
higher up in the spectrum are delocalized. Exactly as in the case of
Anderson-type models, the localized and delocalized modes are separated by a
critical point in the spectrum, the so called mobility edge
\cite{Kovacs:2012zq}. The transition at the mobility edge turned out to be a
genuine second order transition characterized by a correlation length critical
exponent compatible with that of the three dimensional Anderson model of the
same symmetry class \cite{Giordano:2013taa}.

When the temperature is lowered towards the chiral transition, the mobility
edge within the spectrum moves down towards the origin. At the temperature
where the mobility edge vanishes, all the localized modes disappear from the
spectrum and even the lowest part of the spectrum becomes delocalized. We note
that this disappearance of localized modes at some temperature is expected, as
we know that at zero temperature the whole spectrum consists of delocalized
modes. However, what is remarkable is that this localization transition occurs
in the crossover temperature region. We emphasize that on the one hand, the
vanishing of the mobility edge happens at a precisely defined temperature. On
the other hand, the chiral transition is only a cross-over and no precise
critical temperature can be associated with it. Therefore in QCD the
question of whether the chiral and the localization transition happen at
exactly the same temperature, is not meaningful.

There are, however, models similar to QCD where a genuine finite temperature
chiral or deconfining phase transition does take place. An example is QCD with
staggered quarks on coarse lattices $(N_t=4)$. This system possesses a genuine
phase transition in the thermodynamic limit, as the spatial box size goes to
infinity while the temporal size is kept fixed at $N_t=4$ \cite{Nt4}. Here it
was shown that the disappearance of localized modes at the low-end of the
Dirac spectrum precisely coincides with the chiral phase transition
\cite{Giordano:2016nuu}. However, the chiral phase transition of $N_t=4$
staggered quarks does not survive in the continuum limit.

QCD in the limit of large quark masses, i.e.\ the quenched theory is another
example of a QCD-like system with a genuine deconfining phase transition. In
the present paper we focus on this system, SU(3) Yang-Mills theory and study
how the Anderson-type localization transition is related to the finite
temperature deconfining phase transition. An advantage of this model compared
to the one mentioned above is that in this case the phase transition survives
the continuum limit. The main result of the present paper is that the lowest
eigenmodes of the Dirac operator are found to become localized exactly at the
critical temperature of the deconfining transition.

We show this by simulating SU(3) lattice gauge theory at different
temperatures above the deconfining phase transition. We compute the low part
of the spectrum of the Dirac operator on these gauge configurations.  For the
gauge ensembles at each temperature we find the mobility edge by analyzing the
unfolded level spacing distribution. In this way we compute the mobility edge
as a function of the temperature (or rather, the inverse gauge coupling
$\beta$) and by extrapolation we find the temperature where the mobility edge
vanishes. This is the critical temperature of the localization transition
where localized quark modes completely disappear from the spectrum. Below this
temperature even the lowest Dirac eigenmodes are delocalized. We find that the
gauge coupling where this happens is compatible with the critical coupling of
the deconfining transition. This demonstrates that indeed the localization and
the deconfining transition are strongly related.

Another question we would like to touch upon here concerns the nature of the
localized eigenmodes of the Dirac operator and the possible role instantons
might play in the localization transition. A plausible understanding of the
QCD vacuum holds that low eigenmodes of the Dirac operator are essentially
linear combinations of approximate instanton and antiinstanton zero
modes \cite{Schafer:1996wv}. In this framework the localization transition
might be understood as follows. In the Euclidean formulation of QCD
the temperature is set by the temporal size of the box as 
\be
  T = \frac{1}{aN_t},
\ee
where $N_t$ is the temporal box size in lattice units and $a$ is the
physical lattice spacing. As the temperature is increased above the cross-over
temperature, the temporal box size become too small for instantons of typical
size and they are effectively squeezed out of the system. As a result, above
the cross-over the instanton density drastically decreases and the
corresponding approximate zero modes, i.e.\ the low-end of the Dirac spectrum
is depleted.  Due to the lower instanton density and smaller instanton size
the quarks cease to be able to hop from instanton to instanton and get
localized. It was verified in a numerical simulation that an instanton gas
with realistic QCD parameters is indeed capable of producing such a transition
\cite{GarciaGarcia:2005vj}.

If this picture is valid, the localized Dirac eigenmodes are essentially
approximate instanton zero modes. To resolve instantons, a fine enough grid is
needed, therefore the quenched theory where the phase transition survives in
the continuum limit, is well suited for studying the role of instanton zero
modes. Our results indicate that already at $N_t=6$, above the critical
temperature quark eigenmodes connected to instantons can be separated from the
rest of the spectrum, as they show up as a bump in the spectral density around
zero. We demonstrate that these zero mode related quark eigenmodes account
only for a fraction of the localized modes and that fraction becomes ever
smaller as the temperature increases. This indicates that even in the quenched
theory where instantons are expected to be more abundant than in full QCD,
zero modes do not account for the localized quark modes.

The plan of the paper is as follows. In Section \ref{sec:simulation} we
briefly summarize the details of our lattice simulations. In Section
\ref{sec:medge} we discuss the mobility edge and how it can be computed using
the unfolded level spacing distribution. In Section \ref{sec:localization} we
present our main result, the dependence of the mobility edge on the gauge
coupling. By extrapolation we determine the critical coupling where the
mobility edge vanishes. We perform the analysis at three different lattice
spacings, corresponding to $N_t=4, 6$ and $8$, and find that in all cases
the critical coupling for localization is compatible with the point where the
deconfining transition occurs. In Section \ref{sec:instantons} we show that
instanton-related low modes can be separated in the spectrum but most likely
they cannot explain the localized quark modes found at high
temperatures. Finally, in Section \ref{sec:conclusions} we discuss our
conclusions and some further questions not addressed in the present paper.

\section{\label{sec:simulation} Details of the simulation}

We simulate SU(3) lattice gauge theory with the Wilson plaquette action on
lattices of temporal extent $N_t=4, 6$ and $8$. In all three cases we use
several values of the gauge coupling just above the finite temperature
deconfining transition. To assess finite volume effects, at each gauge
coupling we consider at least two different spatial volumes. The deconfining
transition is of first order and the correlation length does not diverge, but
our experience shows that it increases substantially towards the
transition. Therefore, closer to the transition we need to use larger
volumes. We choose the volumes by the requirement that the physical quantity
we are interested in, the mobility edge, be volume independent within the
statistical errors. We summarize the parameters of our simulations in Table
\ref{tab:parameters}.

\begin{table}
 \begin{ruledtabular}
 \begin{tabular}{clrrr}
 $N_t $ & $\beta$ & $N_s$ & Nconf & Nevs \\ \hline
      4 
        & 5.693  & 32    & 1061   &  600 \\
        &        & 40    & 1192   &  1100 \\ 
        &        & 48    & 2381   &  1500 \\ 
        & 5.694  & 32    &  1715   &  600 \\
        &        & 40    & 1005   &  1100  \\ 
        &        & 48    & 2014   &  1600 \\
        & 5.695 & 32    & 2184    &  650 \\
        &        & 40    & 2012   & 1100  \\ 
        &        & 48    &  2028   & 1300  \\ 
        & 5.696 & 32    & 1073    & 900  \\  
        &        & 40    & 1628   &  1000 \\ 
        & 5.6975 & 32    & 2291    & 600  \\ 
        &        & 40    & 1524   & 1100  \\ 
        &        & 48    &  2000   &  1500 \\ 
        & 5.6985 & 40    &  1973   &  1000 \\
        & 5.7 & 24    &  4139   & 600  \\
        &        & 32    & 4040   &  800 \\
        &        & 40    & 1022   & 1000  \\
        & 5.71 & 24    &  2509   &  300 \\ 
        &        & 32    & 2507   &  450 \\ 
        &        & 40    &  1073   &  1100 \\ 
        & 5.74 & 24    &  2024   &  300 \\ 
        &        & 32    & 2501   &  450 \\ 
        &        & 40    &  2390   &  1100 \\ \hline
      6 & 5.897    & 32    &  532   &  900 \\
        &        & 40    &  1249   &  1000 \\
        &        & 48    & 682   & 1350 \\
        & 5.9    & 32    &  998   &  900 \\
        &        & 40    &  925   &  1000 \\
        &        & 48    & 813   & 1350 \\
        & 5.91   &32    & 1068   & 900 \\
        &        & 40    & 834   & 1000 \\
        &        & 48    & 1088   & 1350 \\
        & 5.92   & 32    & 1822   & 600 \\
        &        & 40    & 960   & 1000 \\
        & 5.93   & 32    & 806   & 900 \\
        &        & 40    & 1050   & 1000 \\
        & 5.94   & 40    & 1092   & 1000 \\
        & 5.95   & 32    & 562   & 600 \\
        &        & 40    & 1276   & 1000 \\
        & 5.96   & 32    & 832   & 1000 \\
        &        & 40    & 1032   & 1000 \\
        & 6.0   & 32    & 1392   & 900 \\
        &        & 40    & 1958   & 1000 \\ \hline
      8 & 6.1    & 48    &  439   & 500 \\
        &        & 56    &  681   & 600 \\    
        &        & 64    &  486   & 400 \\
        & 6.15   & 48    &  781   & 500 \\
        &        & 56    &  698   & 600 \\    
        &        & 64    &  385   & 400 \\
        & 6.18   & 48    &  636   & 500 \\
        &        & 56    &  964   & 600 \\    
        &        & 64    &  384   & 400 \\
        & 6.2    & 48    &  675   & 500 \\
        &        & 56    &  778   & 600 \\    
        &        & 64    &  418   & 400 \\
        & 6.25   & 48    &  758   & 500 \\
        &        & 56    &  652   & 600 \\    
        &        & 64    &  320   & 400 \\
        & 6.3    & 48    &  578   & 500 \\
        &        & 56    &  616   & 600 \\    
        &        & 64    &  452   & 400 \\
 \end{tabular}
 \end{ruledtabular}
\caption{\label{tab:parameters} The parameters of the simulations; the
   temporal size of the lattice, the Wilson plaquette gauge coupling, the
   spatial size of the box, the number of configurations and the number of
   Dirac eigenvalues computed on each gauge configuration.}
\end{table}

We calculate the lowest eigenvalues of the Dirac operator on these gauge
configurations. The lattice Dirac operator we consider is the staggered
operator on two times stout smeared gauge links with smearing parameter
$\rho=0.15$ \cite{Aoki:2005vt}. The spectrum of the staggered operator (at
zero quark mass that we consider here) is purely imaginary and symmetric
around zero. Therefore, we will only compute the positive eigenvalues. The
number of eigenvalues we calculate on each configuration is set by the
requirement to capture all the localized modes and also the transition region
from localized to delocalized modes. This is needed for a precise
determination of the mobility edge on each ensemble.

\section{\label{sec:medge} Calculation of the mobility edge}

At high temperature the lowest part of the Dirac spectrum consists of
localized eigenmodes, while higher up in the spectrum the eigenmodes are
delocalized. The boundary between localized and delocalized modes is the
mobility edge, $\lambda_c$. In QCD if the temperature is lowered towards the
transition, the mobility edge shifts down towards zero and at a certain
temperature in the cross-over region it reaches zero. At this point localized
modes completely disappear from the spectrum and even the lowest eigenmodes
becomes delocalized. We expect similar behavior in the quenched theory. Our
main goal is to determine the temperature at which the mobility edge hits zero
and all Dirac modes become delocalized. 

To this end we calculate the mobility edge as a function of the temperature
and by extrapolation find the temperature at which $\lambda_c(T)=0$. Since
strictly speaking in the quenched theory it is not possible to set a physical
scale, we actually compute the mobility edge as a function of the Wilson gauge
coupling $\beta$ and find the gauge coupling where $\lambda_c(\beta)$ becomes
zero. This can be compared to the known critical gauge coupling
$\beta_c\textsuperscript{deconf}$, where the deconfining transition takes
place.

\begin{figure}
\begin{center}
\includegraphics[width=0.95\columnwidth,keepaspectratio]{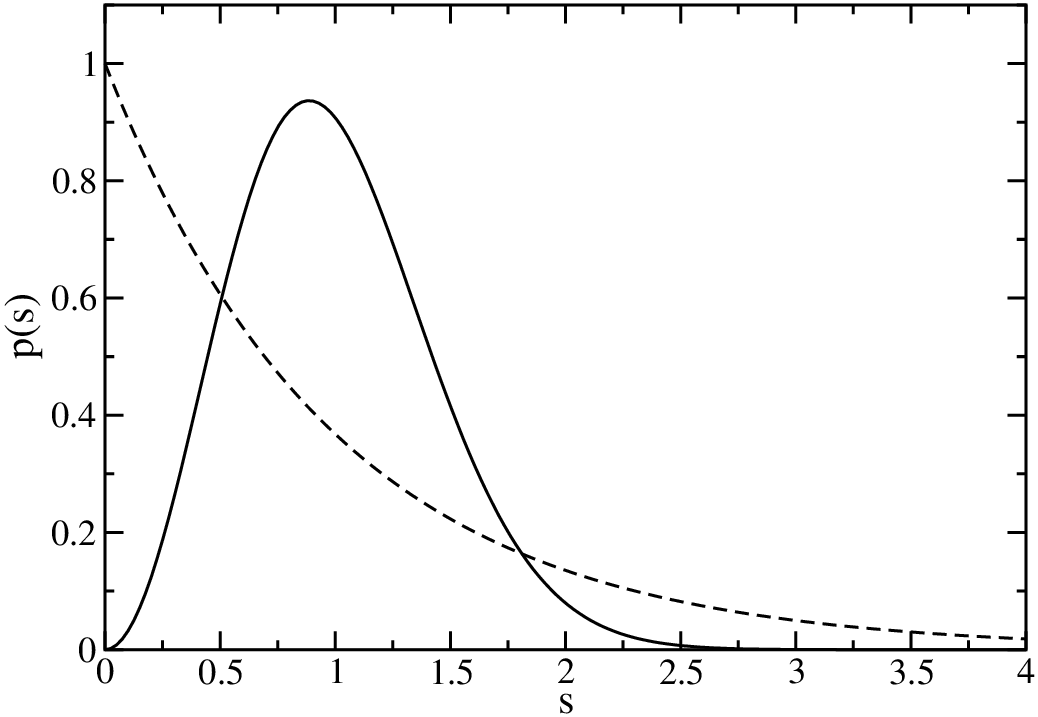}
\caption{\label{fig:lsd2} The exponential distribution (dashed line) of level
  spacings for localized eigenmodes and the unitary Wigner surmise of
  Eq.\ (\ref{eq:wigner}) (solid line) describing the level spacing
  distribution of delocalized modes.}
\end{center}
\end{figure}

On an ensemble of gauge configurations obtained at a given gauge coupling
$\beta$ we use the following procedure to determine the mobility edge. The
simplest way to tell whether the eigenmodes are localized or not is to
consider the unfolded level spacing distribution. After unfolding, which is a
rescaling of the level spacings by their local average (local in the
spectrum), the resulting distribution is known to be universal. In the case of
localized modes it is the exponential distribution $p(s)=\exp(-s)$ following
from an uncorrelated spectrum.  In contrast, the level spacing distribution of
delocalized modes can be accurately described by the unitary Wigner surmise
\be
P\msub{u}(s) = \dfrac{32}{\pi^{2}}s^{2}\cdot
                       \exp\left(-\dfrac{4}{\pi}s^{2}\right)
  \label{eq:wigner}
\ee
corresponding to the symmetry class of the Dirac operator at hand
\cite{Verbaarschot:2000dy} (see Fig.\ \ref{fig:lsd2}). 

\begin{figure}
\begin{center}
\includegraphics[width=0.95\columnwidth,keepaspectratio]{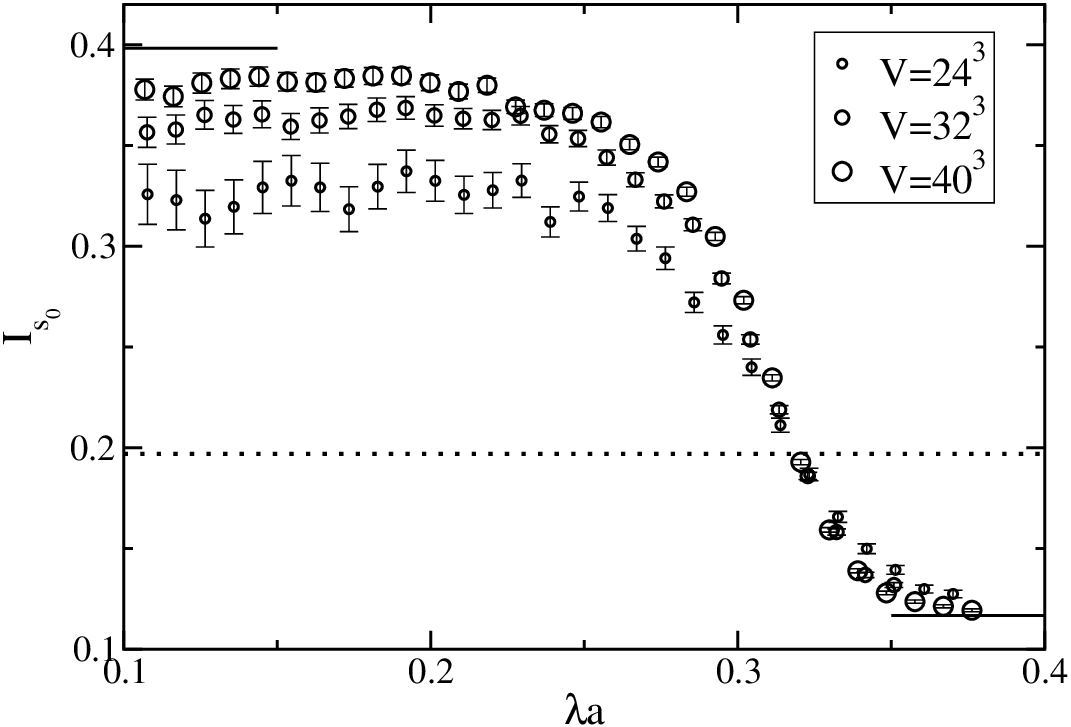}
\caption{\label{fig:is0} The quantity $\is(\lambda)$ as a function of the
  location in the spectrum. $\is$ shows how the level spacing distribution
  changes from exponential to Wigner surmise as the eigenmodes change from
  localized to delocalized across the spectrum. The solid line segments in the
  upper left and lower right corners indicate the limiting values for the
  exponential and the Wigner-Dyson distributions. The dotted line corresponds
  to the critical value of $\is$. Data is shown for $N_t=4, \beta=5.74$ in
  three different lattice volumes with the larger symbols corresponding to
  larger volumes.}
\end{center}
\end{figure}

Since we would like to detect a transition from localized to delocalized modes
within the spectrum, the spectral statistics have to be computed locally. We
do this by subdividing the spectrum into small intervals of width $\Delta
\lambda$ and considering the level spacing distribution separately in each
subinterval as a function of its location in the spectrum. As we move up in
the spectrum and the eigenmodes change from localized to delocalized, the
exponential distribution of level spacings continuously deforms into the
Wigner surmise. To follow this more quantitatively it is convenient to
consider a single quantity characterizing the continuously deforming
distribution. We choose
\be
  \is = \int_0^{s_0} p(s) \, d\!s,
\ee
where the upper limit of the integration was taken to be $s_0=0.508$, the
point where the exponential and the Wigner surmise probability densities first
intersect. This choice was made to maximize the difference in this parameter
between the two limiting distributions. In Fig.\ \ref{fig:is0} we show a
typical plot of how this quantity changes across the spectrum interpolating
from the value corresponding to the exponential distribution to the Wigner
surmise. In the Figure we show $\is$ for different volumes. It is apparent
that as the volume increases the cross-over from localized to delocalized
modes becomes sharper. In fact, in QCD a finite-size scaling study revealed
that in the infinite volume limit the transition is of second order
\cite{Giordano:2013taa}.

Our main concern here is the determination of the critical point $\lambda_c$
in the spectrum where the transition from localized to delocalized modes
happens in the infinite volume limit. We identify the critical point with the
location in the spectrum where $\is$ reaches the value $I_{s_0 cr}=0.196$
corresponding to the critical statistics between the Poisson and Wigner-Dyson
statistics \cite{Shapiro:1993}. Since in the thermodynamic limit the
transition becomes infinitely sharp, instead of $I_{s_0 cr}$, any other value
between the one corresponding to the Poisson and the Wigner-Dyson statistics
could have been chosen. However, our choice turns out to have small finite
size corrections. Another advantage of our choice is that around the point
$I_{s_0 cr}$ the curve $\is(\lambda)$ has an inflection point and it is well
approximated by a straight line. Therefore in the vicinity of this point a
linear fit to the data can be used to obtain a precise determination of the
solution to the equation $\is(\lambda)=I_{s_0 cr}$.

We did this for several choices of $\Delta \lambda$, the width of the spectral
windows where $\is$ was computed. Since the spectral statistics changes along
the spectrum, $\Delta \lambda$ had to be small enough to have this change
negligible within this spectral window. Besides this, there is a trade-off in
the choice of $\Delta \lambda$; narrower spectral windows yield more data
points, however each data point has a larger statistical error, due to the
smaller number of eigenvalues falling in each spectral window and as a result,
a smaller statistics. We chose different values of $\Delta \lambda$ and found
that within reasonable bounds $0.0005 \leq a\, \Delta \lambda \leq 0.005$ the
resulting variation in $\lambda_c$ was well below the statistical
uncertainty. The latter was estimated using the jackknife method.

This procedure yielded $\lambda_c$ for each lattice ensemble. To keep finite
volume corrections under control, we increased the spatial linear size of the
lattice in steps of $8$ until the resulting $\lambda_c$ did not change any
more to within the statistical errors. Not surprisingly, closer to the
deconfining temperature larger volumes were needed for that.

\section{\label{sec:localization} The critical temperature of localization}

\begin{figure}
\begin{center}
\includegraphics[width=0.95\columnwidth,keepaspectratio]{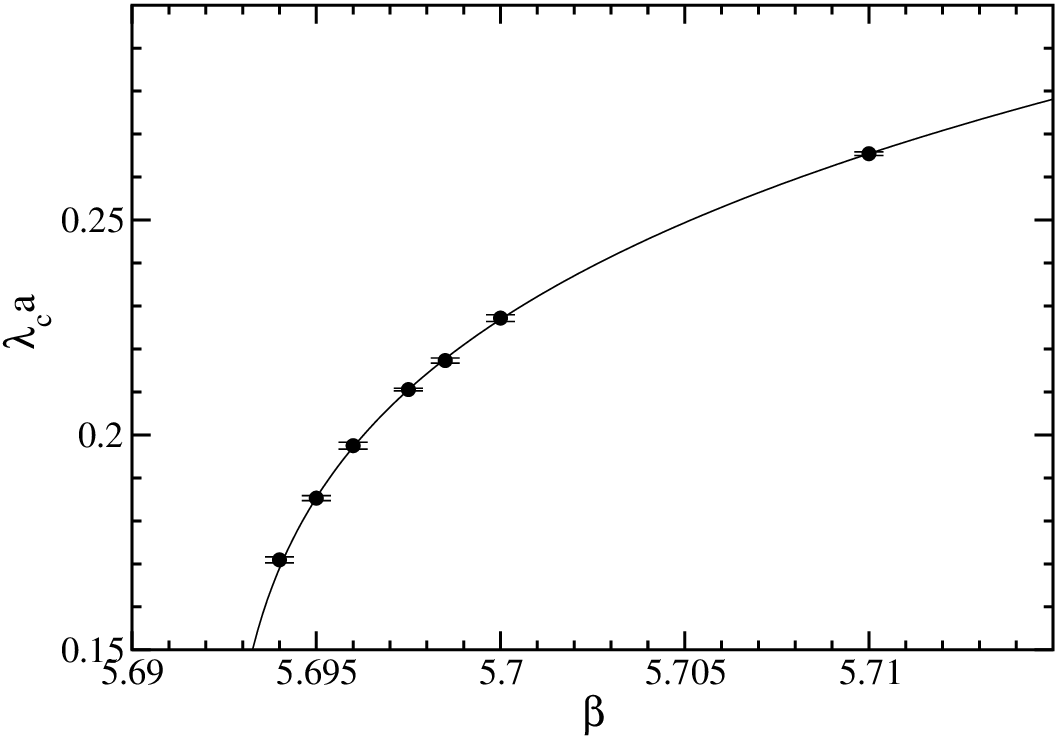}
\caption{\label{fig:lcvsbeta} The mobility edge $ \lambda_c a$ (in lattice
  units) as a function of the inverse gauge coupling for temporal lattice size
  $N_t=4$. The solid line is a fit to the data using the function in
  Eq.\ (\ref{eq:fitfunc}). }
\end{center}
\end{figure}

To determine the dependence of the mobility edge, $\lambda_c$, on the inverse
gauge coupling $\beta$, we repeated the above procedure for different values
of $\beta$ above the deconfining temperature. As an illustration, in
Fig.\ \ref{fig:lcvsbeta} we show $\lambda_c$ versus $\beta$ for the lattices
of temporal size $N_t=4$. The analogous graphs for other values of $N_t$ are
qualitatively similar. In order to determine $\beta_{c}\textsuperscript{loc}$,
the critical coupling for the localization transition, we fitted the
parameters $b,c$ and $\beta_c$ of the function

\be
   a\, \lambda_c(\beta) = c\, (\beta - \beta_{c}\textsuperscript{loc})^b
    \label{eq:fitfunc}
\ee
to the data points\footnote{To avoid the complications and ambiguity coming
  from setting the scale, throughout the analysis we use the dimensionless
  quantity $a \lambda_c$. This is perfectly legitimate since we are only
  interested in the critical point $\beta_{c}\textsuperscript{loc}$ where
  $\lambda_c$ vanishes}. The fit range was chosen by two criteria. The
lower end was restricted because close to the deconfining transition larger
volumes were needed and here we included only those points where simulations
could be done on large enough volumes to ensure no finite volume corrections
to $\lambda_c$. The upper end of the fit range was limited by the fact that
the chosen functional form of $\lambda_c(\beta)$ had a finite range of
validity. At the upper end we included as many data points as was possible to
keep the quality of the fit (measured by the $\chi^2$) acceptable. In the
Figure we show only the points that were used for the fit. Since all the data
points used originated from independent simulations, performed at different
values of the gauge coupling, they are also statistically independent and the
uncertainty of the fit parameters could be easily estimated.

In Table \ref{tab:betac} we give the main result of the paper, the critical
gauge coupling, where localized eigenmodes disappear from the spectrum. In
comparison we also quote the critical coupling for the deconfining transition
which is known from the literature \cite{Francis:2015lha}. Clearly the
critical couplings for the deconfining and the localization transition
coincide to within the numerical uncertainties. Moreover, this happens for
$N_t=4, 6$ and $8$, i.e.\ at three different lattice spacings. We would like
to point out that according to present-day standards of QCD thermodynamics
simulations, even $N_t=8$ lattices around $T_c$ are not considered to be very
fine lattices. However, since in the quenched theory the transition happens at
a relatively higher temperature than in full QCD, these lattices are not very
coarse. Indeed, at the critical point for $N_t=8$ the lattice spacing set by
the Sommer scale \cite{Sommer:1993ce}, $r_0$ is $a=0.085$~fm. Our $N_t=8$
simulations are performed on even finer lattices with
$0.07$~fm$<a<0.08$~fm. It is thus very likely that the same behavior persists
in the continuum limit. This provides strong evidence that the deconfining and
the localization transition are linked.


\bgroup
\def\arraystretch{3}
\begin{table}
\begin{tabular}{llllll}
\hline
\hline
 $N_t$ &\centering $\beta_{c}$\textsuperscript{deconf} &
\centering$\beta_{c}$\textsuperscript{loc} & b & c & fit range
   \tabularnewline   \hline
4 & 5.69254(24)  & 5.69245(17)  & 0.1861(6) & 0.563(2)  & 5.695-5.71 \\ \hline
6 & 5.8941(5)    & 5.8935(16)   & 0.1580(8) & 0.320(1)  & 5.91-5.96  \\ \hline
8 & 6.0624(10)   & 6.057(4)     & 0.164(4)  & 0.233(2)  & 6.08-6.18  \\ 
\hline
\hline     
 \end{tabular}
\caption{\label{tab:betac} The critical gauge couplings where the deconfining
  and the localization transition occur for temporal lattice sizes of $N_t=4,
  6$ and $8$. We also list the fit parameters $b$ and $c$ appearing in
  Eq.\ (\ref{eq:fitfunc}), as well as the best fit range in $\beta$, selected by
  the criteria described in the text.}
\end{table}
\egroup

\section{\label{sec:instantons} Localized modes and instantons}

In the present Section we would like to study the possible relationship
between instanton-antiinstanton zero modes and localized modes. The
simplest question that arises in this connection is how the number of
localized modes compares with that of the topology related eigenmodes. Since
we computed the mobility edge for each ensemble, the average number of
localized modes per configuration, i.e.\ the number of eigenvalues below the
mobility edge can be easily counted. 

A more difficult question is how to separate the topology related modes from
the rest of the spectrum. In fact, above the deconfining transition where the
instantons form a dilute gas, this is possible by looking at the spectral
density of a Dirac operator with sufficiently good chiral properties. Indeed,
it was found that the eigenvalues corresponding to approximate instanton and
antiinstanton zero modes show up in the spectrum of the overlap Dirac operator
as a bump in the spectral density close to zero \cite{Edwards:1999zm}. We
emphasize that these eigenmodes are not the zero modes corresponding to the
net topological charge, but small complex eigenvalues of the overlap Dirac
operator produced by a dilute gas of instantons and antiinstantons. More
recently a very detailed study of this part of the spectrum appeared and it
was found that this feature in the spectrum persists even if dynamical fermions
are present \cite{Alexandru:2015fxa}. 

\begin{figure}
\begin{center}
\includegraphics[width=0.95\columnwidth,keepaspectratio]{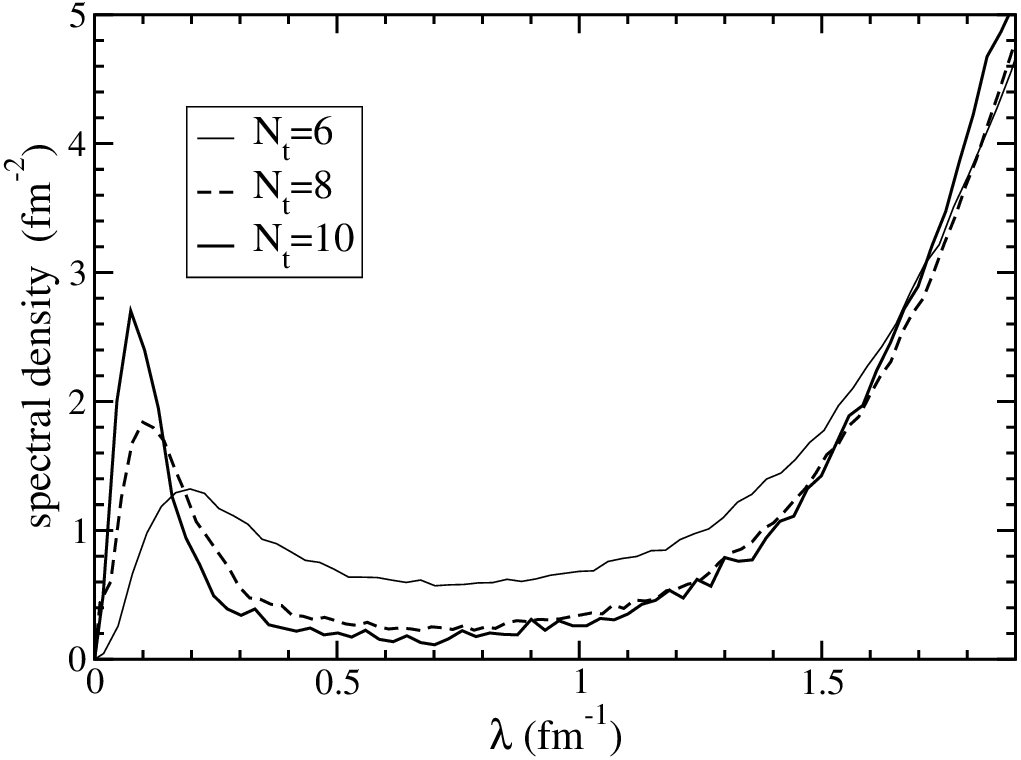}
\caption{\label{fig:spdens} The spectral density of the Dirac operator at a
  temperature slightly above the deconfining phase transition $(T=1.05
  T_c)$. The data sets represent data for different lattice spacings
  corresponding to temporal lattice sizes of $N_t=6,8$ and $10$. The density
  is understood to be normalized with the three-dimensional (spatial) volume.}
\end{center}
\end{figure}

In Fig.\ \ref{fig:spdens} we show the spectral density of the Dirac operator
used in the present paper on $N_t=6,8$ and 10 lattices just above the
deconfining transition at $T=1.05T_c$. We set the scale using the Sommer
parameter, $r_0=0.49$~fm. Already for $N_t=6$ the instanton related modes
appear to be separated from the bulk of the spectrum and this separation
becomes even more pronounced on the finer lattices. We note here that in
Ref.\ \cite{Edwards:1999zm} this spectral structure around zero was seen only
with the overlap operator but not in the spectrum of the staggered
operator. However, the lattices we use here are finer, and in contrast to
Ref.\ \cite{Edwards:1999zm}, we use an improved staggered operator with two
times stout smeared gauge links. Apparently the finer lattices and the
improved staggered operator make it possible to resolve zero modes and
separate them from the bulk of the spectrum \footnote{There is, however, a
  small difference here compared to Ref.\ \cite{Edwards:1999zm}. Since we use
  the staggered Dirac operator, there are no exact zero modes, therefore it is
  not possible to separate them, they also appear in the bump of the spectral
  density.}. In what follows we consider those modes to be instanton related
that are below the minimum between the bump at the origin and the bulk of the
spectrum. 

\begin{figure}
\begin{center}
\includegraphics[width=0.95\columnwidth,keepaspectratio]{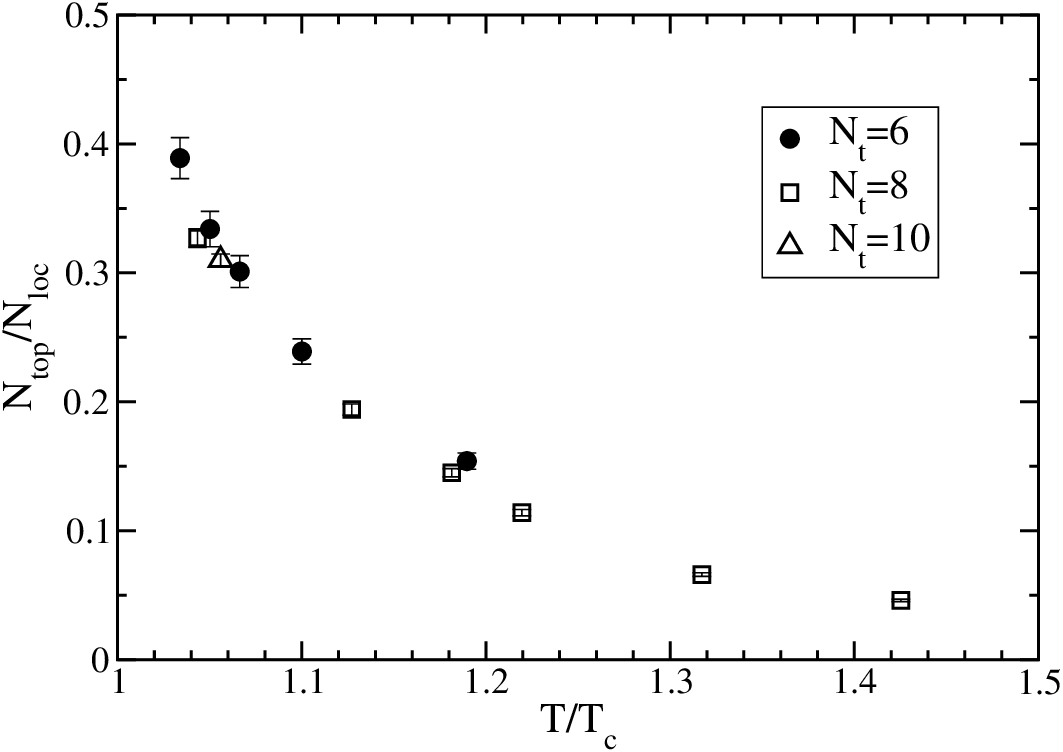}
\caption{\label{fig:topmodes} The ratio of the number of instanton zero mode
  related small eigenvalues to the total number of localized modes as a
  function of the temperature normalized by the critical temperature. The data
  sets correspond to lattices with different lattice spacings, namely
  temporal lattice sizes of $N_t=6,8$ and one point for $N_t=10$. 
}
\end{center}
\end{figure}

Now we can compare the number of topology related small modes and the total
number of localized eigenmodes. In Fig.\ \ref{fig:topmodes}, as a function of
the temperature in units of $T_c$, we plot the fraction of localized modes
accounted for by instantons and antiinstantons on the $N_t=6$ and $N_t=8$
lattices. In the same figure we also included a data point on a finer,
$N_t=10$ lattice. By extrapolating the data down to the critical temperature,
we estimate that just above the deconfining transition about 60\% of the
localized modes can be identified as topology related near zero modes. As the
temperature increases, this fraction falls rapidly. Clearly, above the
deconfining transition most of the localized low modes cannot be identified
with instanton related close to zero modes. This is especially true farther
above the critical temperature. Within our numerical accuracy the data appears
to be scaling which indicates that the statement that topology related modes
do not account for localized modes, is also true in the continuum limit. 

\section{\label{sec:conclusions} Conclusions}

In the present paper we studied the localization transition in SU(3)
Yang-Mills theory. We determined the temperature dependence of the mobility
edge, separating localized and delocalized eigenmodes in the Dirac
spectrum. By extrapolating the mobility edge as a function of the gauge
coupling, we calculated the temperature where the mobility edge reaches
zero. At this temperature localized modes completely disappear from the
spectrum and for smaller temperatures even the lowest Dirac eigenmodes are
delocalized. We found that the temperature where this occurs precisely agrees
with the critical temperature of the deconfining phase transition. This
indicates that the deconfining and the localization transition are strongly
related phenomena. It would be also interesting to understand the details of
the physical mechanism of the why these two phenomena are related. 

We also identified the part of the Dirac spectrum consisting of close to zero
modes produced by instantons and antiinstantons. Just above the transition
they account for a bit more than half of the localized modes. However, the
fraction of instanton zero mode related localized modes falls sharply with
increasing temperature. This comes about from a combination of several
effects. On the one hand, with increasing temperature, instantons are squeezed
out of the system and their density falls rapidly. On the other hand, the
number of localized modes decreases much slower since the decreasing of the
spectral density is somewhat compensated by the fact that the mobility edge
moves up with increasing temperature. To summarize, even in the quenched
theory where the instanton density is higher than in QCD, most of the
localized modes cannot be understood as approximate zero modes. It still
remains to be seen whether localized modes can be associated to some simply
identifiable structure in the gauge field. A first step in this direction
might be the finding that localized modes are spatially correlated with local
fluctuations in the Polyakov loop \cite{Bruckmann:2011cc} and the topological
charge \cite{Cossu:2016scb}. Recently it has been found that just above $T_c$
the structure of gauge field configurations might be more complicated than a
dilute gas of non-interacting instantons \cite{Faccioli:2013ja}. The possible
connection between these recently found dyon structures and localized
Dirac eigenmodes is an interesting possibility still to be explored.

\section*{Acknowledgments}

We thank Matteo Giordano for useful discussions, correspondence and a careful
reading of the manuscript, Ferenc Pittler for his help with the code and Guido
Cossu and Ivan Horvath for their comments and discussions. We acknowledge
support from the Hungarian Science Fund under grant number OTKA-K-113034.


\end{document}